**Effect of demagnetization factor dependence on energy of ultra-thin ferromagnetic films with four layers**

P. Samarasekara

Department of Physics, University of Peradeniya, Peradeniya, Sri Lanka

**Abstract**

Simple cubic and body centered cubic ferromagnetic lattices with four layers were studied using Heisenberg Hamiltonian. According to 3-D plots, the films with four layers can be easily oriented in certain directions under the influence of particular demagnetization factor and angles for both sc(001) and bcc(001) ferromagnetic lattice structures. A flat part can be seen in the middle of 3-D plots in addition to periodic variations. When the demagnetization factor is given by $\frac{N_d}{\mu_0 \omega}=6$, sc(001) film with four layers can be easily oriented in 0.6 radians direction for the energy parameter values used in this simulation. Under the influence of demagnetization factor given by $\frac{N_d}{\mu_0 \omega}=5.2$, thin film of bcc(001) lattice with four layers can be easily oriented along 0.63 radians direction.

**1. Introduction:**

For the first time, the effect of demagnetization factor on energy of ferromagnetic thin films with four layers will be described in this report. Earlier the energy of ultra-thin ferromagnetic films with two and three layers has been studied by us [1]. The film with two layers was equivalent to an oriented film, when anisotropy constants do not vary inside the film. But the energy of films with three layers indicates periodic variation [1]. Introducing second order perturbation induces some sudden overshooting of energy curves, compared with smooth energy curves obtained for oriented ferromagnetic ultra thin films [1]. The energy of oriented thick ferromagnetic films with 10000 layers has been studied [2]. Also the thick films have been studied using Heisenberg Hamiltonian with 2nd order perturbation [3].

Studies of exchange anisotropy have received a wide attention in last decade, because of the difficulties of physical understanding of exchange anisotropy and to its application in magnetic media technology and magnetic sensors [4]. Magnetic properties of ferromagnetic thin films and multi layers have been extensively investigated because of their potential impact on magnetic recording devices. The magnetic properties of thin films of ferromagnetic materials have been investigated using the Bloch spin-wave theory earlier [5]. The magnetization of some thin films shows an in plane orientation due to the dipole interaction. Due to the broken symmetry uniaxial anisotropy energies at the surfaces of the film, the perpendicular magnetization is preferential. But due to the strain induced distortion in the inner layers, bulk anisotropy energies will appear absent or very small in the ideal crystal. Some thin films indicate a tetragonal distortion resulting in stress-induced uniaxial anisotropy energy in the inner layers with perpendicular orientation of easy axis. The magnetic in-plane anisotropy of a square two-dimensional (2D) Heisenberg ferromagnet in the presence of magnetic dipole interaction has been determined earlier [6]. The long range character of the dipole interaction itself is sufficient to stabilize the magnetization in 2-D magnet. Also the easy and hard axes of the magnetization with respect to lattice frame are determined by the anisotropies. Magnetic properties of the Ising ferromagnetic thin films with alternating superlattice layers were investigated [7]. In addition to these, Monte Carlo simulations of hysteresis loops of ferromagnetic thin films have been theoretically traced [8].

According to our previous studies, the stress induced anisotropy plays a vital role in ferrite thin films [10, 12]. The unperturbed, 2$^{nd}$ order perturbed and 3$^{rd}$ order perturbed energy of spinel ferrite and ferromagnetic films was determined by us [11, 13, 14, 15, 16, 17, 18].

## 2. Model:

The classical model of Heisenberg Hamiltonian of any ferromagnetic film can be generally represented by following equation [1, 2].

$$H = -\frac{J}{2}\sum_{m,n}\vec{S}_m \cdot \vec{S}_n + \frac{\omega}{2}\sum_{m \neq n}\left(\frac{\vec{S}_m \cdot \vec{S}_n}{r_{mn}^3} - \frac{3(\vec{S}_m \cdot \vec{r}_{mn})(\vec{r}_{mn} \cdot \vec{S}_n)}{r_{mn}^5}\right) - \sum_m D_{\lambda_m}^{(2)}(S_m^z)^2 - \sum_m D_{\lambda_m}^{(4)}(S_m^z)^4$$

$$-\sum_{m,n}[\vec{H} - (N_d \vec{S}_n / \mu_0)] \cdot \vec{S}_m - \sum_m K_s \sin 2\theta_m$$

The total energy is given by following equation [1].

$$E(\theta) = E_0 + \vec{\alpha} \cdot \vec{\varepsilon} + \frac{1}{2}\vec{\varepsilon} \cdot C \cdot \vec{\varepsilon} = E_0 - \frac{1}{2}\vec{\alpha} \cdot C^+ \cdot \vec{\alpha} \qquad \textbf{(A)}$$

Matrix elements of above matrix C are given by

$$C_{mn} = -(JZ_{|m-n|} - \frac{\omega}{4}\Phi_{|m-n|}) - \frac{3\omega}{4}\cos 2\theta \Phi_{|m-n|} + \frac{2N_d}{\mu_0}$$

$$+ \delta_{mn}\{\sum_{\lambda=1}^{N}[JZ_{|m-\lambda|} - \Phi_{|m-\lambda|}(\frac{\omega}{4} + \frac{3\omega}{4}\cos 2\theta)] - 2(\sin^2\theta - \cos^2\theta)D_m^{(2)}$$

$$+ 4\cos^2\theta(\cos^2\theta - 3\sin^2\theta)D_m^{(4)} + H_{in}\sin\theta + H_{out}\cos\theta - \frac{4N_d}{\mu_0} + 4K_s \sin 2\theta\} \quad (1)$$

$\vec{\alpha}(\varepsilon) = \vec{B}(\theta)\sin 2\theta$ are the terms of matrices with

$$B_\lambda(\theta) = -\frac{3\omega}{4}\sum_{m=1}^{N}\Phi_{|\lambda-m|} + D_\lambda^{(2)} + 2D_\lambda^{(4)}\cos^2\theta \quad (2)$$

Here [2]

$$E_0 = -\frac{J}{2}[NZ_0 + 2(N-1)Z_1] + \{N\Phi_0 + 2(N-1)\Phi_1\}(\frac{\omega}{8} + \frac{3\omega}{8}\cos 2\theta)$$

$$- N(\cos^2\theta D_m^{(2)} + \cos^4\theta D_m^{(4)} + H_{in}\sin\theta + H_{out}\cos\theta - \frac{N_d}{\mu_0} + K_s \sin 2\theta)$$

$E_0$ is the energy of the oriented thin ferromagnetic film. Here $J, Z_{|m-n|}$, $\omega$, $\Phi_{|m-n|}$, $\theta$, $D_m^{(2)}, D_m^{(4)}, H_{in}, H_{out}, N_d, K_s$, m, n and N are spin exchange interaction, number of nearest spin neighbors, strength of long range dipole interaction, constants for partial summation of dipole interaction, azimuthal angle of spin, second and fourth order anisotropy constants, in plane and out of plane applied magnetic fields, demagnetization factor, stress induced anisotropy constant, spin plane indices and total number of layers in film, respectively. When the stress applies normal to the film plane, the angle between m$^{th}$ spin and the stress is $\theta_m$.

For most ferromagnetic films, $Z_{\delta\geq 2} = \Phi_{\delta\geq 2} = 0$. If the anisotropy constants do not vary within the film, then $D_m^{(2)}$ or $D_m^{(4)}$ is constant for any layer.

From equation number 1, the matrix elements of matrix C can be given as following.

$$C_{11}=C_{44}= JZ_1 - \frac{\omega}{4}\Phi_1(1+3\cos 2\theta) - 2(\sin^2\theta-\cos^2\theta)D_m^{(2)}$$

$$+ 4\cos^2\theta(\cos^2\theta-3\sin^2\theta)D_m^{(4)} + H_{in}\sin\theta + H_{out}\cos\theta - \frac{2N_d}{\mu_0} + 4K_s\sin 2\theta$$

When the difference between two indices (m, n) is 1 or -1,

$$C_{12}=C_{23}=C_{34}=C_{21}=C_{32}=C_{43}= -JZ_1 + \frac{\omega}{4}\Phi_1(1-3\cos 2\theta) + \frac{2N_d}{\mu_0}$$

$$C_{22}=C_{33}= 2JZ_1 - \frac{\omega}{2}\Phi_1(1+3\cos 2\theta) - 2(\sin^2\theta-\cos^2\theta)D_m^{(2)}$$

$$+ 4\cos^2\theta(\cos^2\theta-3\sin^2\theta)D_m^{(4)} + H_{in}\sin\theta + H_{out}\cos\theta - \frac{2N_d}{\mu_0} + 4K_s\sin 2\theta$$

From equation 2, $B_1(\theta) = B_4(\theta) = -\frac{3\omega}{4}(\Phi_0+\Phi_1) + D_m^{(2)} + 2D_m^{(4)}\cos^2\theta$

$B_2(\theta) = B_3(\theta) = -\frac{3\omega}{4}(\Phi_0+2\Phi_1) + D_m^{(2)} + 2D_m^{(4)}\cos^2\theta$

Therefore, $C_{11}=C_{44}$, $C_{22}=C_{33}$, $C_{21}=C_{12}=C_{23}=C_{32}=C_{34}=C_{43}$

This simulation will be carried out for

$$\frac{J}{\omega} = \frac{D_m^{(2)}}{\omega} = \frac{H_{in}}{\omega} = \frac{H_{out}}{\omega} = \frac{K_s}{\omega} = 10 \text{ and } \frac{D_m^{(4)}}{\omega} = 5$$

For sc(001) lattice, $Z_0=4$, $Z_1=1$, $Z_2=0$, $\Phi_0=9.0336$, $\Phi_1= -0.3275$ and $\Phi_2=0$ [9],

$$\frac{C_{11}}{\omega} = \frac{C_{44}}{\omega} = 10.08 + 20.2456\cos 2\theta$$

$$+ 20\cos^2\theta(\cos^2\theta-3\sin^2\theta) - \frac{2N_d}{\mu_0\omega} + 10\cos\theta + 10\sin\theta + 40\sin 2\theta$$

$$\frac{C_{22}}{\omega} = \frac{C_{33}}{\omega} = 20.164 + 20.49\cos 2\theta$$

$$+ 20\cos^2\theta(\cos^2\theta-3\sin^2\theta) - \frac{2N_d}{\mu_0\omega} + 10\cos\theta + 10\sin\theta + 40\sin 2\theta$$

$$\frac{C_{12}}{\omega} = \frac{C_{21}}{\omega} = \frac{C_{23}}{\omega} = \frac{C_{32}}{\omega} = \frac{C_{34}}{\omega} = \frac{C_{43}}{\omega} = -10.08 + 0.2456\cos 2\theta + \frac{2N_d}{\mu_0\omega}$$

$$\frac{\alpha_1}{\omega} = \frac{\alpha_4}{\omega} = (3.47 + 10\cos^2\theta)\sin 2\theta$$

$$\frac{\alpha_2}{\omega} = \frac{\alpha_3}{\omega} = (3.716 + 10\cos^2\theta)\sin 2\theta$$

$$\frac{E_0}{\omega} = -105.73 + 12.81\cos 2\theta$$

$$- 4(10\cos^2\theta + 5\cos^4\theta + 10\cos\theta + 10\sin\theta - \frac{N_d}{\mu_0\omega} + 10\sin 2\theta)$$

$$\frac{C_{13}}{\omega} = \frac{C_{14}}{\omega} = \frac{C_{24}}{\omega} = \frac{C_{31}}{\omega} = \frac{C_{41}}{\omega} = \frac{C_{42}}{\omega} = \frac{2N_d}{\mu_0\omega}$$

Under some special conditions [1], $C^+$ is the standard inverse of matrix C, given by matrix element $C^+_{mn} = \frac{cofactor C_{nm}}{\det C}$. For the convenience, the matrix elements $C^+_{mn}$ will be given in terms of $C_{11}$, $C_{22}$, $C_{12}$ and $C_{13}$ only.

Determinant of $C = C_{11}[C_{22}(C_{22}C_{11}-C_{12}^2)-C_{12}(C_{12}C_{11}-C_{12}C_{13})-C_{13}(C_{12}^2-C_{22}C_{13})]$

$-C_{12}[C_{12}(C_{22}C_{11}-C_{12}^2)-C_{12}(C_{13}C_{11}-C_{12}C_{13})+C_{13}(C_{13}C_{12}-C_{22}C_{13})]$

$+C_{13}[C_{12}(C_{12}C_{11}-C_{12}C_{13})-C_{22}(C_{13}C_{11}-C_{12}C_{13})+C_{13}(C_{13}^2-C_{12}C_{13})]$

$-C_{13}[C_{12}(C_{12}^2-C_{22}C_{13})-C_{22}(C_{13}C_{12}-C_{13}C_{22})+C_{12}(C_{13}^2-C_{12}C_{13})]$

$$C_{11}^+ = C_{44}^+ = \frac{C_{22}(C_{22}C_{11} - C_{12}^2) - C_{12}(C_{12}C_{11} - C_{12}C_{13}) + C_{13}(C_{12}^2 - C_{22}C_{13})}{\det C}$$

$$C_{12}^+ = C_{21}^+ = -\frac{C_{12}(C_{22}C_{11} - C_{12}^2) - C_{13}(C_{12}C_{11} - C_{12}C_{13}) + C_{13}(C_{12}^2 - C_{22}C_{13})}{\det C}$$

$$C_{23}^+ = C_{32}^+ = -\frac{C_{11}(C_{12}C_{11} - C_{12}C_{13}) - C_{13}(C_{12}C_{11} - C_{13}^2) + C_{13}(C_{12}^2 - C_{12}C_{13})}{\det C}$$

$$C_{13}^+ = C_{31}^+ = \frac{C_{12}(C_{12}C_{11} - C_{13}C_{12}) - C_{13}(C_{22}C_{11} - C_{13}^2) + C_{13}(C_{12}C_{22} - C_{12}C_{13})}{\det C}$$

$$C_{22}^+ = C_{33}^+ = \frac{C_{11}(C_{22}C_{11} - C_{12}^2) - C_{13}(C_{13}C_{11} - C_{12}C_{13}) + C_{13}(C_{12}C_{13} - C_{22}C_{13})}{\det C}$$

$$C_{14}^+ = C_{41}^+ = -\frac{C_{12}(C_{12}^2 - C_{13}C_{22}) - C_{13}(C_{22}C_{12} - C_{12}C_{13}) + C_{13}(C_{22}^2 - C_{12}^2)}{\det C}$$

$$C_{24}^+ = C_{42}^+ = \frac{C_{11}(C_{12}^2 - C_{13}C_{22}) - C_{13}(C_{12}^2 - C_{13}^2) + C_{13}(C_{12}C_{22} - C_{12}C_{13})}{\det C}$$

$$C_{34}^+ = C_{43}^+ = -\frac{C_{11}(C_{12}C_{22} - C_{12}C_{13}) - C_{12}(C_{12}^2 - C_{13}C_{12}) + C_{13}(C_{12}C_{13} - C_{22}C_{13})}{\det C}$$

From equation (A),

$$E(\theta) = E_0 - \alpha_1^2(C_{11}^+ + C_{14}^+) - \alpha_1\alpha_2(C_{12}^+ + C_{13}^+ + C_{24}^+ + C_{34}^+) - \alpha_2^2(C_{22}^+ + C_{23}^+) \quad (3)$$

## 3. Results and discussion:

Matrix elements of inverse matrix $C_{mn}^+$ can be found from above equations, and hence total energy can be found from above equation 3. Then 3-D plot of energy versus angle and $\frac{N_d}{\mu_0 \omega}$ can be given as shown in figure 1. A flat portion can be seen in the middle of the graph. Several energy minimums can be observed at different values of $\frac{N_d}{\mu_0 \omega}$ and angle, indicating that the sc(001) ferromagnetic films with four layers can be easily oriented in these directions. For example, the film can be easily oriented at $\frac{N_d}{\mu_0 \omega} = 6$ in certain directions. The easy directions corresponding to this $\frac{N_d}{\mu_0 \omega}$ can be found from figure 2. According to this graph, this film can be easily oriented in direction given by 0.6 radians.

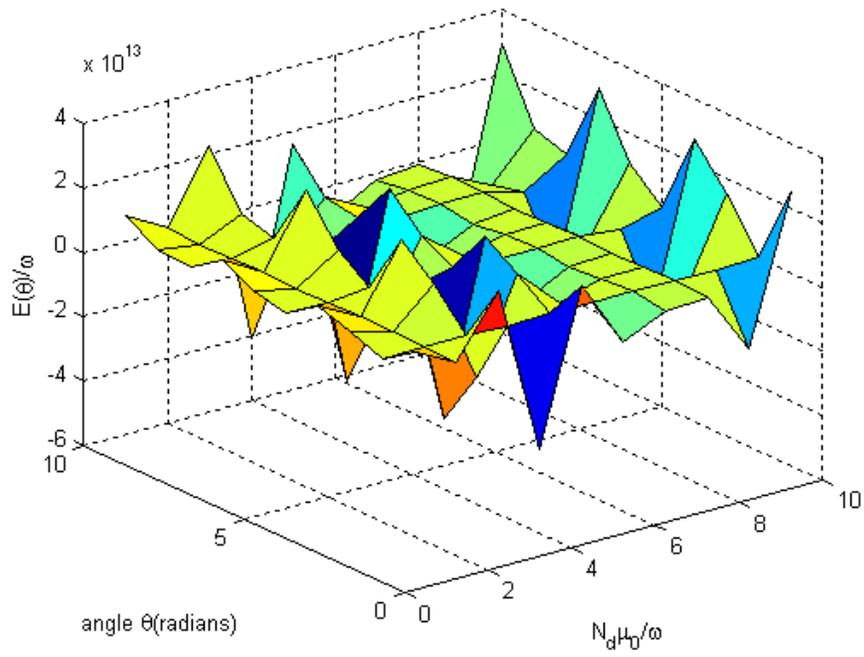

Figure 1: 3-D plot of energy versus angle and $\dfrac{N_d}{\mu_0 \omega}$ for sc(001) lattice

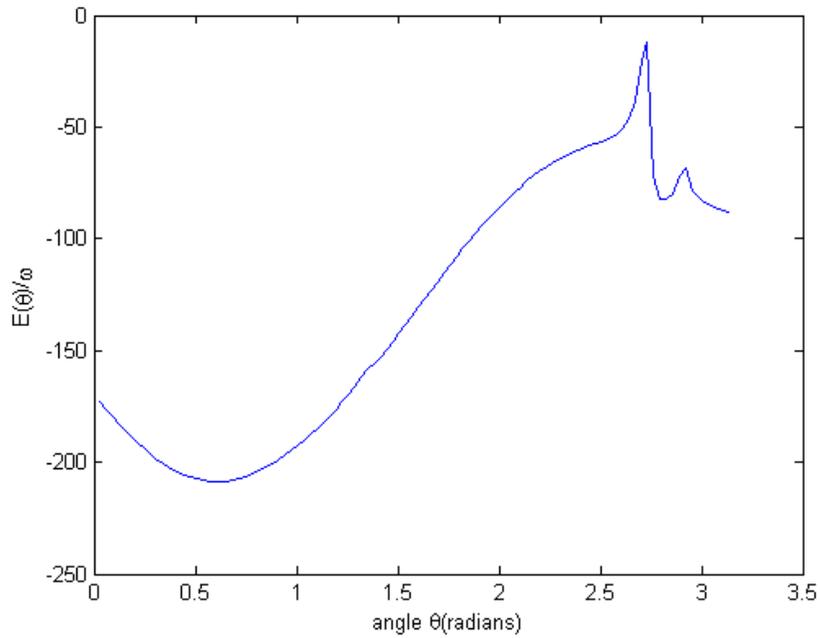

Figure 2: Plot of energy versus angle at $\dfrac{N_d}{\mu_0 \omega}=6$ for sc(001) lattice

For bcc(001) lattice, $Z_0=0$, $Z_1=4$, $Z_2=0$, $\Phi_0=5.8675$ and $\Phi_1=2.7126$ [9],

$$\frac{C_{12}}{\omega} = \frac{C_{21}}{\omega} = \frac{C_{23}}{\omega} = \frac{C_{32}}{\omega} = \frac{C_{34}}{\omega} = \frac{C_{43}}{\omega} = -39.32 - 2.03\cos 2\theta + \frac{2N_d}{\mu_0\omega}$$

$$\frac{C_{13}}{\omega} = \frac{C_{31}}{\omega} = \frac{C_{14}}{\omega} = \frac{C_{24}}{\omega} = \frac{C_{41}}{\omega} = \frac{C_{42}}{\omega} = \frac{2N_d}{\mu_0\omega}$$

$$\frac{C_{11}}{\omega} = \frac{C_{44}}{\omega} = 39.32 - 2.03\cos 2\theta - \frac{2N_d}{\mu_0\omega} + 20\cos^2\theta(\cos^2\theta - 3\sin^2\theta)$$

$$+ 10\sin\theta + 10\cos\theta + 40\sin 2\theta$$

$$\frac{C_{22}}{\omega} = \frac{C_{33}}{\omega} = 78.64 - 4.07\cos 2\theta - \frac{2N_d}{\mu_0\omega} + 20\cos^2\theta(\cos^2\theta - 3\sin^2\theta)$$

$$+ 10\sin\theta + 10\cos\theta + 40\sin 2\theta$$

$$\frac{\alpha_1}{\omega} = \frac{\alpha_4}{\omega} = (3.565 + 10\cos^2\theta)\sin 2\theta$$

$$\frac{\alpha_2}{\omega} = \frac{\alpha_3}{\omega} = (1.53 + 10\cos^2\theta)\sin 2\theta$$

$$\frac{E_0}{\omega} = -115 + 14.9\cos 2\theta$$

$$- 4(10\cos^2\theta + 5\cos^4\theta + 10\sin\theta + 10\cos\theta - \frac{N_d}{\mu_0\omega} + 10\sin 2\theta)$$

3-D plot of energy versus angle and $\frac{N_d}{\mu_0\omega}$ for bcc(001) lattice is given in figure 3. Although a flat portion can be observed at the middle of this graph, the shape of this graph is slightly different from graph 1. Energy is minimum at some certain values of angles and $\frac{N_d}{\mu_0\omega}$ indicating that film can be easily oriented along those directions under the influence of certain demagnetization factors. At $\frac{N_d}{\mu_0\omega}$=5.2, film can be easily oriented in certain directions. The graph between energy and angle was drawn in order to determine these easy directions at $\frac{N_d}{\mu_0\omega}$=5.2 as shown in figure 4. This film can be easily oriented in 0.63 radians direction.

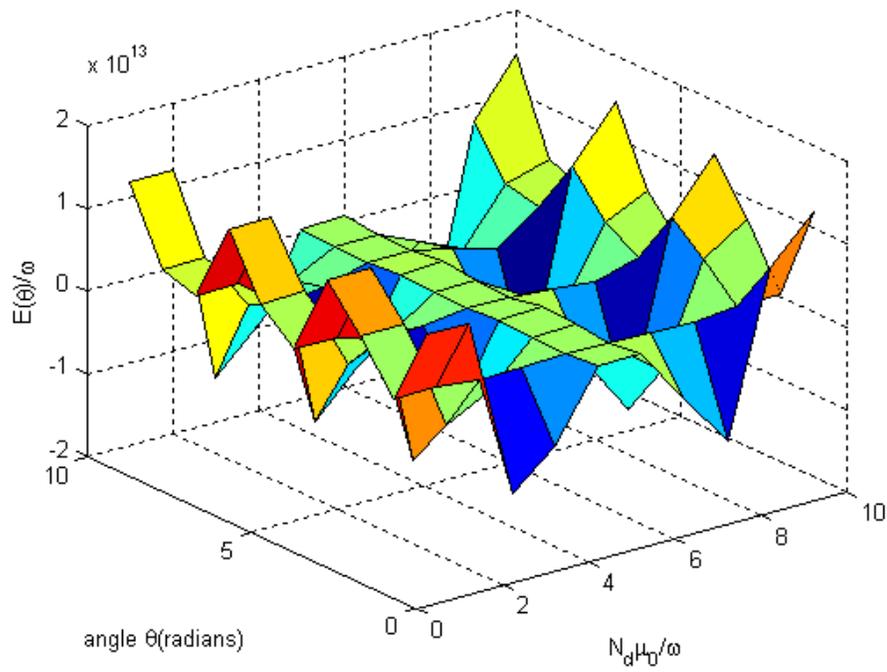

Figure 3: 3-D plot of energy versus angle and $\dfrac{N_d}{\mu_0 \omega}$ for bcc(001) lattice

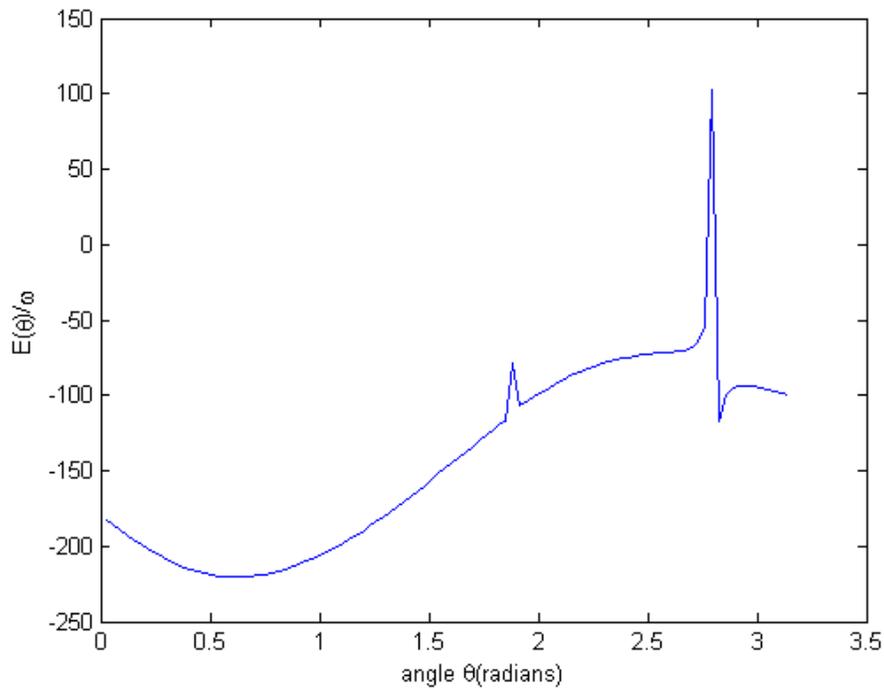

Figure 4: Plot of energy versus angle at $\dfrac{N_d}{\mu_0 \omega}=5.2$ for bcc(001) lattice

## 4. Conclusion:

For both sc(001) and bcc(001) ferromagnetic lattice structures, energy is minimum at certain values of demagnetization factor and angles. Thin film of sc(001) lattice with four layers can be easily oriented along 0.6 radians direction under the influence of demagnetization factor given by $\frac{N_d}{\mu_0 \omega}=6$. When the demagnetization factor is given by $\frac{N_d}{\mu_0 \omega}=5.2$, bcc(001) film with four layers can be easily oriented in 0.63 radians direction. Although this simulation was carried out for $\frac{J}{\omega}=\frac{D_m^{(2)}}{\omega}=\frac{H_{in}}{\omega}=\frac{H_{out}}{\omega}=\frac{K_s}{\omega}=10$ and $\frac{D_m^{(4)}}{\omega}=5$, this same simulation can be carried out for any value of $\frac{J}{\omega}, \frac{D_m^{(2)}}{\omega}, \frac{H_{in}}{\omega}, \frac{H_{out}}{\omega}, \frac{K_s}{\omega}$ and $\frac{D_m^{(4)}}{\omega}$.